\documentstyle[12pt]{article}
\begin{document}
\setlength{\unitlength}{1mm}

\title{\bf  Ground-state instabilities in
the one-dimensional Penson-Kolb-Hubbard model}
\author{A. Belkasri and  F.D. Buzatu \thanks{E-mail:
BUZATU@theor.jinrc.dubna.su}
 \ \thanks{Permanent address:{ \it Department of Theoretical Physics,
 Institute for Physics and
 Nuclear Engineering, Institute for Atomic Physics, P.O.Box Mg-6, R-76900
 M\u{a}gurele, Bucharest, Romania}.} \\
 \it Bogoliubov Laboratory of Theoretical Physics \\
 \it Joint Institute for Nuclear Research  \\
 \it Dubna, Moscow Region, 141980 Russia }

\date{}
\maketitle

\begin{center}
\vspace{0.5cm}
Preprint Dubna E17-95-373 \\
Submitted to {\em Physical Review B}
\end{center}

\vspace{1cm}
\begin{abstract}
Different kinds of  instabilities ($CDW$, $SDW$, $SS$) in the 1D Hubbard
model with pair-hopping interaction are investigated using an approximate
Bethe-Salpeter equation. The study is performed at any density of electrons
and for arbitrary values of the model parameters. In the absence
of the on-site interaction, no transition occurs at half filling for
any finite negative pair-hopping parameter, in agreement with recent
results; our calculation suggests that the Penson-Kolb model ($t,W$)
with $\mid W/t\mid<\pi/\sin k^{}_F$
behaves qualitatively like the Hubbard model ($t,U$).
Phase diagrams of the Penson-Kolb-Hubbard model ($t,U,W$)
at various densities are presented.

\end{abstract}
\newpage

\section{Introduction}
\hspace{\parindent}The studies of the one-dimensional
models of correlated electronic systems can
be a primary source to understand the occurence of the high temperature
superconductivity in materials which physics is mainly two-dimensional.
In these high-$T_{c}$ superconductors the ``Cooper pairs'' are extremely
small with coherence lengths comparable with the size of the unit cell.
Various mechanisms can lead to this local pairing \cite{Mic}.
In this paper we consider a model which can be relevant to the high-$T_{c}$
superconductivity because it contains not only such a local pairing but
also an on-site electron-electron repulsion, interaction which may lead
to the insulating phase of the cuprates.

The Hamiltonian of the
Penson-Kolb-Hubbard (PKH) model is \cite{Hui}
\begin{equation}
 \begin{array}{ll}
{\cal H}=
-t\sum\limits_{i,\sigma}^{}\left( c^{\dag}_{i+1,\sigma}c^{}_{i,\sigma}
                                  +H.c.\right)
 &+\ U\sum\limits_{i}^{}n^{}_{i,{\scriptscriptstyle \uparrow}}
  n^{}_{i,{\scriptscriptstyle \downarrow}} \\
 &   \\
 &+\ W\sum\limits_{i}^{}\left( c^{\dag}_{i,\scriptscriptstyle \uparrow}
                    c^{\dag}_{i,\scriptscriptstyle \downarrow}
                    c^{}_{i+1,\scriptscriptstyle \downarrow}
                    c^{}_{i+1,\scriptscriptstyle \uparrow}+H.c.\right)
 \end{array}
\end{equation}
where we have used the standard notation for fermion operators :
$c^{\dag}_{i,\sigma}( c^{}_{i,\sigma})$ creates (destroys) an electron
of spin $\sigma=\uparrow,\downarrow$ on the lattice site $i$ and
$n_{i,\sigma}=c^{\dag}_{i,\sigma}c^{}_{i,\sigma}$. In the absence of
the on-site interaction term $U$ , the hamiltonian (1) reduces to the
Penson-Kolb (PK) model \cite{Penson} where the competition between
the single- and pair-hopping of electrons can lead to interesting effects
in 1D; a spin-gap transition at half filling for $W<0$ has been the
subject of some controversy \cite{Hui,Penson,Affleck,Sikkema}.
Recently, some variantes of the PKH model were solved by Bethe ansatz
method \cite{Bedurftig,Bariev1}, where the single-particle hopping term
is modified to include interaction effects: the jumping of an electron
to an empty site differs from that corresponding to an occupied one
\cite{Bariev2}. But the integrability of such model is possible only
under some restrictions on the interaction parameters. Consequently,
it remains interesting to have results depending on all parameters
of the model and in a wide range. That is why we will consider
both positive and negative values for $U$ and $W$ ($t>0$) in Eq. (1) and
arbitrary density; however, our results for the ground-state phase diagram
are valid in a definite range of parameters determined below.

We investigate the possible occurence of instabilities in the ground-state
of the PKH model in the same manner \cite{Buzatu} as it was done for the
1D ($t,U,X$) model \cite{Painelli} : within the (zero-temperature)
Green-function formalism in the Bloch representation, the instabilities
are signaled by the poles of the vertex function $\Gamma $ which obeys
the Bethe-Salpeter equation. We solve this equation in the approximation
when the irreducible vertex part is just the bare potential and the
single-particle propagator has the `free' expression. The imaginary part
of the poles (in the total frequency variable), interpreted as the
inverse of the relaxation time to a new ground-state, gives us the regions
in the parameters space where the instabilities can occur; in the regions
common to more instabilities we choose that phase with the shortest
relaxation time.

\section{Bethe-Salpeter equation}

\hspace{\parindent}To understand  what kinds of instabilities can occur
in the system,
one can investigate the generalized susceptibilities. It is assumed that
we start
from a phase where there is no order parameter, and study the density-density
fluctuations. When the generalized susceptibility is singular,
it is an indication that a spontaneous distortion or ordering can occur
in the system.

The general form of the susceptibility is
\begin{equation}
\chi(k,\omega)=-i\int\! dt\ e^{i\omega t}_{}\
<T\{ {\cal D}(k,t){\cal D}^{\dag}(k,0)\} >
\end{equation}
where $T$ is the usual chronological operator and
${\cal D}(k,t)$
is a density operator. The brackets indicate an average in the
ground-state (for zero-temperature case) or a statistical average
(for finite-temperature case). If
${\cal D}(k,t)$  is a charge-density operator, the corresponding
susceptibily $\Gamma^{}_{CDW}$
will test the {\em charge density wave} ($CDW$)
instability; in a similar way can be defined the susceptibilities
relevant to the other kinds of instabilities:
$\Gamma^{}_{SDW}$
for {\em spin density wave} ($SDW$),
$\Gamma^{}_{SS}$
for {\em singlet superconductivity} ($SS$) and
$\Gamma^{}_{TS}$ for
{\em triplet superconductivity} ($TS$). The correspondence between
these quantities and the components of the vertex function can be
found in Refs. \cite{Buzatu}.

The generalized susceptibilities are two-particle Green functions
and they obey the Bethe-Salpeter equation. In the simplest
approximation, where only the bare interaction is
considered for the irreducible vertex part and
the one-particle propagator $G$ is replaced by the free one $G^{0}$,
it can be written as:
\begin{equation}
\!\!\!\xi \Gamma (k,k^{\prime };K ,\Omega )=\frac{i}{2\pi }
V(k,k^{\prime };K )\!+\!
   \sum\limits_{k^{\prime \prime }}^{} V(k,k^{\prime \prime };K )
   {\cal G}(k^{\prime \prime };K ,\Omega )\Gamma (k^{\prime \prime },
   k^{\prime };K ,\Omega )
\end{equation}
where $\Gamma $ can be any of the quantities $\Gamma^{}_{CDW}$,
$\Gamma^{}_{SDW}$ or $\Gamma^{}_{SS}$ ;
the $TS$ case does not occur in this approximation because the interaction
in the Hamiltonian (1) is only between electrons with opposite spin.
$K$ ($\Omega $) denotes
the transfer momentum (frequency) in the particle-hole
($ph$) channel and the  total momentum (frequency) in the particle-particle
($pp$) channel,

\begin{equation}
\xi=\left\{\begin{array}{rc}
            -1     & CDW      \\
             1     & SDW, SS
            \end{array}
    \right.
\end{equation}

\begin{equation}
 \begin{array}{l}
\hspace{-0.5cm}{\cal G}(k;K ,\Omega)\ =  \\
\hspace{-0.5cm}{\displaystyle \frac{i}{2\pi}}\int\limits_{-\infty}^{+\infty}
\!d\omega \, G^{0}(k+K/2 ,\omega +\Omega /2 )\!\times \!
                       \left\{ \begin{array}{lc}
    \!\!\!G^{0}(k-K/2 ,\omega -\Omega /2 ) \!\!\!   &
                CDW, SDW \\
               &          \\
    \!\!\!G^{0}(K/2-k,\Omega /2 -\omega ) \!\!\!   &
                  SS
                               \end{array}
                       \right.
 \end{array}
\end{equation}
where the addition or subtraction of the $k$-vectors are defined
modulo $2\pi $ (the lattice constant is considered one).
$V$ in Eq. (3) comes from the interaction part of
the Hamiltonian (1) in the Bloch representation and has the expresion
\begin{equation}
V(k,k';K)=\frac{1}{N}\left\{\begin{array}{rc}
            U+2W\cos (k+k')     & CDW, SDW      \\
             U+2W\cos (K)     & SS
            \end{array}
    \right.
\end{equation}
where $N$ denotes the number of sites in the chain.

For the $ph$ channel, Eq. (3) admits a solution of the form
\begin{equation}
\Gamma =\frac{i}{2\pi N}
\widehat{E}^{T}(k)\widehat{X}(K,\Omega )\widehat{E}(k')
\end{equation}
with
\begin{equation}
\widehat{E}(k)=\left( \begin{array}{c}
            1 \\
            \cos k \\
            \sin k
            \end{array} \right) \ \ ,\ \
\widehat{X}(K,\Omega )=\left( \begin{array}{lll}
                              X_{11} & X_{12} & X_{13}   \\
                              X_{21} & X_{22} & X_{23}   \\
                              X_{31} & X_{32} & X_{33}
                              \end{array} \right)
\end{equation}
where $\widehat{E}^{T}$ means the transposed matrix of $\widehat{E}$.
The unknown coefficients $X_{ij}(K,\Omega)$
are determined from the following algebraic system
\begin{equation}
\widehat{M} \widehat{X}
 =\left( \begin{array}{ccc}
      U   & 0 & 0 \\
      0  &  -2W   & 0  \\
      0   &  0   & -2W
         \end{array}
  \right)
\end{equation}
where $\widehat{M}$ is the $3\times 3$ matrix
\begin{equation}
\widehat{M} \equiv \left( \begin{array}{ccc}
  \xi -gU   &   -c_1 U      & -s_1 U    \\
   -2 c_1 W     &    \xi -2 c_2 W  & -2 p W \\
   2 s_1 W      &     2 p W      &  \xi +2 s_2 W

                    \end{array}
             \right)
\end{equation}
and
\begin{equation} \left\{ \begin{array}{l}
g={\displaystyle \frac{1}{N}}\sum\limits^{}_{q}{\cal G}(q;K,\Omega)  \\
  \\
c_n={\displaystyle \frac{1}{N}}\sum\limits_{q}^{}
    \left( \cos q \right)^{n}{\cal G}(q;K ,\Omega) \ ,\ n=1,2 \\
  \\
s_n={\displaystyle \frac{1}{N}}\sum\limits_{q}^{}
    \left( \sin q \right)^{n}{\cal G}(q;K ,\Omega) \ ,\ n=1,2 \\
  \\
p={\displaystyle \frac{1}{N}}\sum\limits_{q}^{}
   \sin q \cos q {\cal G}(q;K ,\Omega)
   \end{array} \right.
\end{equation}
Since for the $SS$ case $V(k,k';K)\equiv V(K)$, the solution of the Eq. (3)
in the $pp$ channel reads immediatly
\begin{equation}
\Gamma^{}_{SS}=\frac{i}{2\pi N}
\frac{V(K)}{1-V(K)g(K,\Omega )}
\end{equation}

An instability in the ground-state of the system occurs
when the determinant $D$ of the $\widehat M$ matrix vanishes for the
first time starting from the noninteracting case,
indicating that the $X_{ij}$ diverge and consequently $\Gamma$
diverges. Following Refs. \cite{Buzatu}, we look for the
$\Gamma $-poles of the form
\begin{equation}
\Omega =E_{exc}+iT  \ \ , \ \
E_{exc}=\left\{\begin{array}{ccc}
            0                  & CDW, SDW  & (K=2k^{}_{F})  \\
            2\varepsilon^{}_{F}   & SS        & (K=0)
            \end{array}
    \right.
\end{equation}
where $E_{exc}$ is the excitation energy to provide the system
to undergo a phase transition; $T$ is the inverse of the relaxation
time of the unstable ground-state and can be also regarded as
a `temperature'; $k^{}_F =\pi n/2$ is the Fermi momentum
($n$ being the density of electrons)
and $\varepsilon^{}_F =-2t\cos k^{}_F$.
In this case, the determinant $D$ has the form
\begin{equation}
D\simeq\mu +\rho^{}_{F}\lambda \ln \left| \frac{\Omega_{0}}{T} \right|
\label{det}
\end{equation}
which is valid for $\mid T/\Omega_0\mid \ll 1$.
This condition is similar to
the BCS theory, where only the excitations of electrons around
the Fermi level (with energies much less than
the Debye energy) are taken into account. We will use the expression
(\ref{det})
to find the solutions of the equation $D=0$ for $|T/\Omega_0|<1$ and we
expect the results to be reliable at least for not too big values of
$|\lambda /\mu |$ which plays the role of the coupling constant.

The parameters in Eq.(\ref{det}) are given by
\begin{equation}
\mu =\left\{ \begin{array}{ll}
  {\displaystyle \pm 1 + {2\over\pi}\left(\sin k^{}_F \right) w
  \pm{{w^2}\over{\pi^2}} \pm{1\over {2\pi^2}}
  {{k_F^2-\ln^2 |\cos k^{}_F |}\over {\sin^2 k^{}_F}} u w+ }& \\
{\displaystyle \frac{1}{2\pi^{3}\sin k^{}_{F}} \left(
\frac{2k^{}_{F}}{\tan k^{}_{F}}
\ln |\cos k^{}_F|
 + k^{2}_F-\ln^2 |\cos k^{}_F | \right) u w^2} &
                  \begin{array}{l}
                        SDW \\
                        CDW
                  \end{array}                  \\
                                 &  \\
  1   & SS
            \end{array}
     \right.
\end{equation}

\begin{equation}
\lambda =\left\{ \begin{array}{ll}
 \!{\displaystyle  -\frac{1}{2}\left[ u + 2 w \pm {2\over \pi}
\left ( \sin k^{}_F + {{\ln |\cos k^{}_F |}\over {\sin k^{}_F}}\right ) u w
\pm {2\over {\pi}}\left( \sin k^{}_F \right) w^2 + \right. }  &   \\
{\displaystyle \left.
 {1\over {\pi^2}}\left( 1-2{{k^{}_F}\over {\tan k^{}_F}}
 +{{k^{2}_F}\over {\sin^2 k^{}_F}} + 2\ln |\cos k^{}_F | \right)
 u w^2 \right]  }
     & \begin{array}{l}
                                               SDW \\
                                               CDW
                                          \end{array}  \\
                            &          \\
 \!{\displaystyle  u + 2 w }  &   SS
                 \end{array}
          \right.
\end{equation}
where $u\equiv U/t$, $w\equiv W/t$ ;
\begin{equation}
\Omega_{0}=8t\sin^{2}k^{}_{F} \left\{ \begin{array}{cc}
                 \left( \cos k^{}_F \right)^{-1}  & CDW,SDW \\
                                         1   &  SS
                                              \end{array}
                                         \right.
\end{equation}

\begin{equation}
\rho^{}_{F}=\left( 2\pi \sin k^{}_F \right)^{-1}
\end{equation}
$\rho^{}_{F}/t$ being the density of states at the Fermi level.

It follows that a transition to an ordered phase will
occur at the critical `temperature'
\begin{equation}
T_{c}=|\Omega_{0}|\exp\left(\frac{\mu}{\rho^{}_{F}\lambda}\right) \ \ ,
\end{equation}
with $\mu/\lambda <0 $ (so that $|T_c/\Omega_o |<1$).
In order to get the
phase diagram for the Hamiltonian (1), we determine at first the
regions in the $(w,u)$-space where the quantity $\mu/\lambda$ is negative
(for each case: $CDW$, $SDW$ or $SS$); when more than one instability can
occur in a given region, we decide for the phase
which is held first, {\em i.e.} with the shortest relaxation time
(or equivalently, with the biggest critical `temperature' $T_c$). We
restrict our considerations to that region of the $(w,u)$-space
containing the origin $u=w=0$ and
where $\lambda /\mu $ never becomes infinite.

\newpage

\section{ Phase diagram of the PK model }

\hspace{\parindent}For $U=0$ the Hamiltonian (1) reduces to the PK
model \cite{Penson}. By comparing the various critical `temperatures'
(for $CDW$, $SDW$ and $SS$) we get the phase diagram plotted in Fig. 1:
the investigated region is $|w|\sin k^{}_{F}<\pi $ following from the
condition
$|T_{c}/\Omega_{0}|<1$,
as discussed above. For $w>0$ we get
only a $SDW$. For $w<0$ there is a $SS$ phase for densities less than
a critical one $n_c>\frac{2}{\pi}Arccos(e^{-2})\simeq 0.914$ and a $CDW$
phase for $n>n_c$; the critical density $n_c$ tends to one as
$n\rightarrow 1$. According to our calculations, at half filling the
system is in a $SDW$ state for $w>0$ and in a $CDW$ phase for $w<0$.
Let us remark that at half filling the condition
$|T_{c}/\Omega_{0}|=1$
(when the effective coupling constant becomes infinite)
determines the limits $w=\pm \pi$. It is interesting to note that
close to our limit $w=-\pi$, around
the value $w\simeq -3.5$, Sikkema and Affleck \cite{Sikkema} found a
phase separation transition; in our approach, the existence of such a
transition can be in principle analysed by calculating the compressibility
in the homogeneus phase \cite{Penc}.

Since at $w=0$ the electrons move freely and at large negative $w$ the
ground-state contains only doubly occupied and empty sites,
it was argued \cite{Penson} that should be a `pairing transition' at
some negative $w_c$. Exact diagonalizations on chains up to $12$ sites
\cite{Hui,Penson} have shown that at half filling this transition occurs
around $w_c\simeq -1.4$; but conformal field methods \cite{Affleck} and
renormalization group studies \cite{Sikkema} have shown that $w_c=0$.
In our calculation the only transition which occurs at half filling
(in the investigated range) is for $w=0$.

\section{Phase diagrams for the PKH model}

\hspace{\parindent}In the limit of half filling ($n\rightarrow 1$)
we found a $SDW-CDW$ transition along the curve
\begin{equation}
u=-2w\left[ 1 - {{w^2}\over  {1+(1-{1/{\pi^2}})w^2}  } \right]
\end{equation}
in agreement with the prediction of Hui and Doniach \cite{Hui} who found
such a transition for $u,|w|\ll 1$ along the line $u\simeq -2w$. The
obtained phase diagrams at various densities are presented in Figs. 2--4.
The first remark is the superconducting phase can not appear when
$u+2w>0$; this is due to the form of the bare potential $V(K)$, given by
Eq. (6) in the $pp$ channel, which becomes repulsive in that region. However,
it follows from Figs. 2--4 that we do not get systematically a $SS$ phase for
$V(K)<0$; some regions of $SDW$ or (and) $CDW$ phase still remain. The
$SS$ region decreases by increasing the density; it disappears for $n=1$.
Let us note that near half filling our phase diagram
is qualitatively in agreement with that obtained by
Hui and Doniach \cite{Hui} who studied the same model
with $u>0$ and $w<0$ using exact
diagonalization for samples of up to $12$ sites. For example,
for rather big values of $u$ and $w$ they found a sequence
$SDW\rightarrow CDW\rightarrow SS$ in passing from $u>-w$ to $u<-w$; this
one can be also observed in our results from Fig. 2.

\section{Conclusions}

\hspace{\parindent}In this paper we have presented phase diagrams
for the PKH model at arbitrary densities of electrons and for moderate
values of the parameters $W/t$ and $U/t$. Our mean-field-type approximation
predicts results consistent with other works done at half filling.
In the particular case of the PK model ($U=0$) we have found for
$|W|/t<\pi/\sin k^{}_{F} $ a phase diagram similar to that
corresponding to the
Hubbard model in the same approximation \cite{Buzatu}; at half filling,
the only transition which occurs is a $SDW-CDW$ at $W=0$.
However,
beyond the limits $|W|/t=\pi /\sin k^{}_F$ indicated by our approach, we
expect a qualitative change in the ground-state of the PK model. To what
extend this fact can be related or not with the phase separation transition
found by Sikkema and Affleck \cite{Sikkema} at $W/t\simeq -3.5$ near
half filling, is a subject for further investigations.

\vspace{1cm}

\section*{Aknowledgments}
The authors are grateful to Prof. N. Plakida for enlightened remarks
related to this subject. One of the authors (A.B) would like to thank
to the Directorate of the J.I.N.R. (Dubna, Russia) for the hospitality
intented to him during the preparation of this work.

\end{document}